\input harvmac.tex
%\draft
\Title{\vbox{\baselineskip12pt\hbox{IP-BBSR/96-67}\hbox{CTP-TAMU-63/96}
\hbox {MRI-PHYS/96/34}
\hbox{hep-th/9701115}
}}
{\vbox{\centerline{Notes on Axion, Inflation and Graceful Exit}\smallskip
\centerline{ in Stringy Cosmology}}}
\centerline{\bf Jnanadeva Maharana$^1$, Sudipta Mukherji$^2$ and  Sudhakar
Panda$^3$}
\smallskip\centerline{$^1$\it Institute of Physics, Bhubaneswar 751 005, 
INDIA} 
\smallskip\centerline{e-mail: maharana@iopb.ernet.in} 
\smallskip\centerline{ Jawaharlal Nehru Fellow}
\smallskip\centerline{$^2$\it Center for Theoretical Physics, Department of 
Physics}
\smallskip\centerline{\it Texas A \& M University, College Station, Texas
77843-4242, USA}
\smallskip\centerline{e-mail: mukherji@bose.tamu.edu}
\smallskip\centerline{$^3$\it Mehta Research Institute of Mathematics and
Mathematical Physics}
\smallskip\centerline{\it10, Kasturba Gandhi Marg, Allahabad 211 002, INDIA}
\smallskip\centerline{e-mail: panda@mri.ernet.in}

\vskip .3in
We study the classical equations of motion and the corresponding 
Wheeler-De Witt equations for tree level string effective action
with the dilaton and axion. The graceful exit problem in certain 
cases is then analysed.

\Date{1/97}
\vfill
\eject

\def \bp {{\bar \phi}}

\def \ta {\tau}

%%%%%%%%
\lref\old{J.Ellis, K.Enqvist, D.V.Nanopolous and M.Quiros, Nucl. Phys.
{\bf B277} (1986) 231; K.Maeda and M.D.Pollock, Phys. Lett. {\bf B173}
(1986) 251; P.Binetruy and M.K. Gaillard, Phys. Rev. {\bf D34} (1986)
3069; D.G. Boulware and S. Deser, Phys. Lett. {\bf B175} (1986)
409; R.C. Myers, Phys. Lett. {\bf B199} (1987) 371; I.
Antoniadis, C. Bachas, J. Ellis and D.V. Nanopoulos, Phys. Lett.
{\bf B211} (1988) 393; Nucl. Phys. {\bf B328} (1989) 117; 
I. Antoniadis and C. Kounnas, Nucl. Phys. {\bf B175} (1987) 729;
S.Kalara, C.Kounnas and K.A. Olive, Phys. Lett. {\bf B215}
(1988) 265; S. Kalara and K.A. Olive, Phys. Lett. {\bf B218}
(1989) 148;
B.A. Campbell, N.Kaloper and K.A. Olive, Phys. Lett. {\bf B277}
(1992) 265; A.A. Tseytlin, Mod. Phys. Lett. {\bf A6} (1991) 1721;
N.Kaloper and K.A. Olive, Astroparticle Phys. {\bf 1} (1993) 185.}

\lref\odd{G. Veneziano, Phys. Lett. {\bf B265} (1991) 287; K.A.
Meissner and G. Veneziano, Mod. Phys. Lett. {\bf A6} (1991) 337;
Phys. Lett. {\bf B267} (1991) 33; M. Gasperini, J. Maharana and
G. Veneziano, Phys. Lett. {\bf B272} (1991) 277; Phys. Lett.
{\bf B296} (1992) 51.}

\lref\gv{M. Gasperini and G. Veneziano, Astroparticle Phys. {\bf 1} 
(1993) 317;
Mod. Phys. Lett. {\bf A8} (1993) 3701; Phys. Rev. {\bf D50} (1994) 2519.}

\lref\veneziano{G. Veneziano, Phys. Lett {\bf B265} (1991) 287.}

\lref\gmv{M. Gasperini, J. Maharana and G. Veneziano, {\it
Graceful Exit in Quantum String Cosmology}, hep-th/9602087.}

\lref\bv{R. Brustein and G. Veneziano, Phys. Lett {\bf B329} (1994) 
429; N.Kaloper, R. Madden and K.A. Olive, Phys. Lett. {\bf B371} (1996)
34; Nucl. Phys. {\bf B452} (1996) 677; R. Easther, K. Maeda and
D. Wands, Phys. Rev. {\bf D53} (1996) 4247.} 

\lref\emw{R. Easther, K. Maeda and D. Wands, {\it Tree-level String 
Cosmology}, hep-th/9509074.}

\lref\callan{C. Callan, D. Friedan, E. Martinec and M. Perry, Nucl. Phys. 
{\bf B262}, (1985), 593.}
\lref\ew{For earlier works see K. Enqvist, S. Mohanty and D.V.
Nanopoulos, Phys. Lett. {\bf B192} (1987) 327; Int. J.Mod. Phys.
{\bf A4} (1989) 873; A. Lyons and S.W. Hawking, Phys. Rev {\bf
D44} (1991) 3902; M.D. Pollock, Int. J.Mod. Phys. {\bf A7}
(1992) 4149; J. Wang, Phys. Rev. {\bf D45} (1992) 412; 
 M.C. Bento and O. Bertolami, Class. Quantum Grav.
{\bf 12} (1995) 1919; A.A. Kehagias and A. Lukas hep-th/9602084; M.
Gasperini and G. Veneziano, hep-th/9602078.}

\lref\clw{E. Copeland, A. Lahiri and D. Wands, Phys. Rev. {\bf D50}
(1994) 4868.}

\lref\pope{S. Mukherji, hep-th/9609048;
H. Lu, S. Mukherji, C. N. Pope and K. W. Xu, hep-th/9610107;
H. Lu, S. Mukherji, C. N. Pope, hep-th/9612224.}

\lref\lsey{J. E. Lidsey, Phys. Rev {\bf D52} (1995) 5407;
gr-qc/9605017; gr-qc/9609063.}

\lref\ovrut{A. Lukas, B. Ovrut and D. Waldram, hep-th/9608195.}
 %%%%%%%%%%%%%%%%%%%%%
%\newsec{Introduction}
Recently cosmological implications of string theory has attracted
considerable attention \refs{\old -\odd}. 
It is expected that string theory will 
provide
answers related to the evolution of the Universe in early epochs and
eventually shed light  on the creation of the Universe. A very attractive
proposal has been put forward in order to provide the mechanism for
inflation and it stems from the scenario of the pre-big-bang cosmology
\refs{\gv} in the string theoretic frame work.
We recall that  the cosmological solution of the tree level string
effective action admits two separate branches labelled as $(+)$ and $(-)$.
There is a solution in the $(+)$ - branch which corresponds to weakly
coupled dilaton for a cold and flat Universe in the $ t \rightarrow -
\infty$ limit. As time increases to zero, this background
configuration evolves towards a strong coupling regime and the solution
admits positive Hubble parameter, H, and rapidly  accelerating expansion
towards strong curvature, progressing to a singularity in its future.
On the other hand, the (-)-branch proceeds to a large spatially flat
Universe such that the Hubble parameter is positive (expanding Universe),
but with deceleration. The singularity is in the past for this particular 
solution.
Furthermore, it can be smoothly joined to a FRW cosmological solution in
the late epoch. Naturally, in view of the existence of such cosmological
solutions, one is tempted to identify the former solution with the
inflationary phase of the Universe whereas the latter endowed with several
attractive features of the late time cosmology.

Indeed, these two solutions are related to one another by the  stringy
symmetry known as scale factor duality (SFD) which is a part of the larger
T-duality symmetry group, $O(d,d)$. If we accept that these two solutions
correspond, in reality, to a single solution in the temporal evolution of
the Universe, we can envision the following scenario. There is a
pole-driven superinflationary phase for $t<0$. It is driven by the dilaton
kinetic energy term and the dilaton potential does not affect the growth
of the scale factor in this regime. Subsequently, for $t >0$, we go over
to the expanding, decelerating Universe. However, one encounters the
problem of graceful exit in string cosmology as is the case in 
any theory which incorporates inflations.
Recently, it has been shown that the graceful exit problem persists\refs{\bv}
for the string theoretic approach to cosmology when
one considers the mechanism for branch changes in the classical frame
work. The no-go theorems have emphasised that the branch changes are not
possible even if we include effects of the axion and perfect fluids in the
string evolution equations in addition to the dilaton potential.
Therefore, so long as one considers tree level string effective action,
the graceful exit problem awaits its resolution at the classical level.

A minisuperspace model for spatially homogeneous Bianchi I Universe has
been considered by Gasperini, Veneziano and one of us (JM) \refs{\gmv}. 
These authors
have shown that the wave function of the Universe, derived from the
Wheeler-De Witt equation \refs{\ew}, can be expanded in terms of
plane waves in the
minisuperspace. It was found that configurations associated with the
pre-big-bang correspond to "right moving" waves, whereas those
corresponding to the post-big-bang backgrounds are identified with the
"left moving" waves. Thus the two waves move in the opposite directions of
the effective spatial coordinates. Moreover, the two branches are related
by SFD and a time reversal symmetry. Furthermore, it was
demonstrated by explicit examples \refs{\gmv}
that transitions between the pre- and post-big-bang domains are possible
if the wave function undergoes spatial reflection in the minisuperspace.

The purpose of this note is to explore further the resolution of the
graceful exit problem and construct explicit solutions to the WDW 
equation for
a more general class of dilaton potential or to take into account the
contributions of the axionic field. First, we shall discuss classical
solutions to the string equations of motion and subsequently look for
solutions of the Wheeler-De Witt equation.

%\newsec{More Solutions}
We consider the low energy string effective action 
 \refs{\callan} of the form 
\eqn\action{S = -{1\over {2\lambda^2}}\int d^4x{\sqrt -g}e^{-\phi}
(R + \partial_\mu\phi\partial_\nu\phi - {1\over {12}} H^2 +V(\phi)),}
where $\phi$ is the dilaton field (with $e^{\phi\over 2}$
being the string coupling constant), $H_{\mu\nu\rho}$ is the 
field strength of the anti-symmetric tensor filed $B_{\mu\nu}$.
There can also be cosmological constant term ($\Lambda$) in the
action, the precise form of which depends on the detail of the 
compactification scheme used to bring the 10 dimensional action
to 4 dimension. Moreover, non-preturbatively there can be dilaton
potential contribution to the action. The contribution of cosmological 
term and the dilaton potential have been grouped together in $V(\phi)$
of the above action.
 
 In the cosmological case, all the background fields only depend on cosmic
time, t and we can write the metric and and the antisymmetric tensor field
in the following form

\eqn\mat{g_{\mu \nu} = \pmatrix{-1&0\cr 0&G_{ij}(t)}}

\eqn\mat{B_{\mu \nu} = \pmatrix{0&0\cr 0&B_{ij}(t)}}

Where $G_{ij}$ and $B_{ij}$  are $3 \times 3$ symmetric and antisymmetric
matrices respectively. The action can be rewritten in a manifestly
$O(3,3)$ form when the backgrounds assume time dependence.

\eqn\oaction{S~=~{-\lambda\over 2} \int dt e^{- \bar \phi} [(\partial_t\bp)^2 + 
{1\over 8}{\rm Tr}( \partial_t M \eta \partial_t M \eta) + V],}
where 
\eqn\barp{\bp = \phi - {1\over 2}{\rm ln}|G_{ij}|.}
Here a constant spatial volume factor has been absorbed into the the
definition of $\bp$ and $M$ is a symmetric $6\times 6$ metrix given by
\eqn\mat{M = \pmatrix{G^{-1}&-G^{-1}B\cr BG^{-1}&G-BG^{-1}B}}
and we have defined the matrices  $G$ and $B$ earlier.
The action is  invariant under $O(3, 3)$ 
\eqn\odd{\bp \rightarrow \bp, ~~~M = \Omega^T M \Omega ,}
where 
\eqn\om{{\Omega^T \eta \Omega} =\eta,~~~\eta = 
\pmatrix{0&I\cr I&0}}
with $M$ satisfying 
\eqn\msatis{M\eta M = \eta .}
Note that the potential $V = V(\bp)$ in order that the action remains
invariant under $O(3,3)$. For diagonal metric and zero torsion, 
with $\Omega = \eta$, this 
transformation reduces to the scale factor duality.
As mentioned earlier, the duality symmetry  
relates  different branches of the scale 
factor and dilaton when we envisage their time evolutions. Of special
interest to us is the one  corresponding  
to acclerated, expanding background 
with growing curvature(pre-big bang) which gets related to another 
with desired late time attributes such as expanding decelerated, 
universe with decreasing curvature (post-big bang).

In what follows, we study the action \action\
and the classical solutions relevant in our context,discuss the 
Wheeler-De Witt equation and analyze the wave function of the universe
when the background is given by spatially flat Robertson-Walker metric with
\eqn\metric{ds^2 = N(t) dt^2 - e^{2\beta(t)\over \sqrt
3}dx_idx_j \delta^{ij}.} 

It is well known that the field strength of the antisymmetric tensor
field, in four dimensions, is related to the pseudoscalar axion through
the Poincare duality transformation
\eqn\psu{H^{\mu\nu\rho} = e^\bp
\epsilon^{\mu\nu\rho\lambda}\partial_\lambda \Theta .}
Furthermore, the equations of motion for $H^{\mu\nu\rho}$ is a
conservation law and thus the equation of motion for $\Theta$ leads to
conservation of the axionic charge. Thus, for the case at hand, the action
can be brought to the following form

\eqn\newac{S = -{\lambda\over 2}\int d\ta\{{1\over N}
(\partial_\tau \bp)^2 -{1\over N} (\partial_\tau\beta)^2 +
{q^2 N\over 2}e^{-2(\sqrt 3\beta + \bp)} + N V(\bp)e^{-2\bp}\}.}
where $dt = e^{-\bp}d\tau$ and $\bp$ has been defined before.
Here the charge q is related to the canonical momentum of $\Theta$, 
 $p_\Theta = {\partial{\cal 
L}\over{\partial\partial_t\Theta}} = q $.

At this stage the following comments are in order. We use the
dilaton time variable, $\tau$, instead of the cosmic time
variable, t, in the definition of the action integral above. This
choice is made for later convenience. If we had used the
integral over cosmic time to define the action integral and
derived the the equations of motion, then the classical
solutions would have taken the known form \refs{\bv} and they
would have satisfied the usual constraint equations. However,
when we choose to work in the $\tau$-variable, the constraint
equation (see below) takes a simpler form; neverthless if we go
over to t-variable we recover the usual constraint equation.
Henceforth, we shall express the equations of motion in terms of
$\tau$ variable and obtain solutions in this variable.

First, we focus our attention to the case where the axionic
charge, $q$, is taken to be zero and we consider the dilaton
potential which is a generalized form of the one considered by the
authors of \refs{\gmv}. In the second example, we shall consider
nonvanishing axionic charge with vanishing dilatonic potential.

When we adopt the first scenario, the equations of motion for 
$\beta$ and the dilaton $\bp$, in the gauge $N=1$ are 
\eqn\eomo{\partial_\tau^2\beta = 0,}

\eqn\eomt{\partial_\tau^2\bp - {1\over 2}{\partial\over{\partial\bp}}
(Ve^{-2\bp}) = 0.}
Furthermore, $\beta$ and $\bp$ must satisfy the following constraint
\eqn\con{(\partial_\tau\bp)^2 - (\partial_\tau\beta)^2 - Ve^{-2\bp} = 0.}

Now with the potential $V = V_0 e^{m\bp},~~~~V_0>0$ and for
$m\neq 2$ the eqn \eomo\ ,\eomt\ 
and \con\ can be solved exactly with the answers
\eqn\sol{\beta =  {k\tau\over \lambda},
~~~\bp ={-2\over (m-2)}{\rm ln}~[{\lambda\sqrt V_0\over k}
{\rm sinh} { k(m-2)\tau\over {2\lambda}}],}
here $k$ is a constant of intergation.
Note that for $m=4$ the above potential is same as in
\refs{\gmv} where the equations of motion could be solved
exactly i.e. $\beta$, thus the scale factor $a$ and $\bp$ could
be expressed in terms of the cosmic time. Thus the present study
is a generalization of \refs{\gmv} for $m\neq 4$.
%and for $V = - V_0 e^{m\bp},~~~~V_0>0$ we get
%\eqn\soll{\beta =  c\tau,~~~\bp ={-2\over (m-2)}{\rm ln}~[{\sqrt V_0\over c}
%cosh { c (m-2)\tau\over 2}]}
%where $c$ is a constant.

However, although we could get $\bp$ and $\beta$ explicitly as
functions of $\tau$, it is 
hard to write them in terms of cosmic time $t$ in closed forms. Neverthless,
it is straight forward to notice that we can get desired solutions
such as the inflationary one in the pre-big-bang regime and
expanding, decelerating one in the positive $\tau$ domain.
From \sol\ we see that for negative $\tau$(or $t$) as $\tau\rightarrow
-\infty$ the scalefactor $a\rightarrow 0$ and we begin in the
weak coupling phase ($\bp\rightarrow -\infty$) and evolve
towards the strong coupling ($\bp\rightarrow +\infty$)
domain as $\tau\rightarrow 0$ from the negative side. 
It is easy to check that in terms of cosmic time, $\tau =0$
corresponds to finite value of $t$. Furthermore, for
$|t|$ less than this value, the fields become complex. 
Hence there would always be a classically forbidden region. In particular,
it is easy to check that close to this region, $\bp$
behaves as 
\eqn\apro{\bp = -{2\over m}{\rm ln}[-
{\lambda V_0\over {k^2}} + {mk^2\over{4\lambda}}t^2 - ({m\over 4} -1)
{mk^4\over {24\lambda^3V_0}}t^4 + .....].}
By using the definition of $H$ and the
defining relation between the cosmic time $t$ and $\tau$ we can
check that $H > 0$ and ${\dot H} >0$ as required by pre
big-bang scenario; here dot denotes derivative with respect to the cosmic time.
 Similarly, it is quite evident that for large
positive $\tau$(or $t$), we proceede towards the scenario of late time
cosmology, $H > 0$, but ${\dot H} < 0$.

If we adopt the Hamiltonian analysis of \refs{\gmv}, then the
the WDW equation in this case can be written down as
\eqn\wdw{[({\partial\over {\partial\bp}})^2 -({\partial\over 
{\partial\beta}})^2 + \lambda^2 V_0 e^{(m-2)\bp}]\Psi(\bp
,\beta) = 0.} 
Note that the appearance of $\lambda^2$ in front of the
potential term can be traced back to the definition of the
action and correspondingly, the canonical momenta are defined
with the factor of $\lambda$. Also, for the background generated
by the chosen potential, the WDW equation \wdw\ can be split
as $\Psi(\bp,\beta) = \psi_k(\bp) e^{-ik\beta}$ where $k$ is a
constant which belongs to the continuous eigenvalue spectrum of 
the momentum operator corresponding to the $\beta$-direction and
the function $\psi_k (\bp)$ satisfies
\eqn\ps{(\partial_\bp^2 +k^2 +\lambda^2 V_0 e^{(m-2)\bp})
\psi_k(\bp) =0,}
whose general solution is a linear combination of first and
second kind Hankel functions, $H_\nu^{(1)}(z)$ and
$H_\nu^{(2)}(z)$ where $\nu ={i2k\over m-2}$ and $z =
{2\lambda\over m-2}{\sqrt V_0} e^{{m-2\over 2}\bp}$. However,
we require the wave-function to be regular in the limit
$\bp\rightarrow \infty$ and this fixes the wave function to be 
\eqn\wsol{\Psi_k(\bp , \beta) = N
H^{(1)}_{i({2k\over{m-2}})}({2\lambda\over {m-2}}{\sqrt V_0} 
e^{{(m-2)\bp\over 2}})e^{-ik\beta},} 
where $N$ is a normalization constant. Thus, asymptotically, 
in the regime $\bp\rightarrow -\infty$, we have
\eqn\wdwlc{\lim_{\bp\rightarrow -\infty} \Psi_k (\bp,\beta) =
\Psi_k^{(+)} - \Psi_k^{(-)},}
where
$$
\Psi_k^{(+)} = i N~ {\rm cosec} ({2ik\over m-2} \pi) e^{{2k\over
m-2}} ({\lambda{\sqrt V_0}\over m-2})^{{2ik\over m-2}}
{e^{-ik(\beta - {2\bp\over m-2})}\over \Gamma ( 1+{2ik\over
m-2})}, 
$$
and
$$
\Psi_k^{(-)}= i N ~{\rm cosec}({2ik\over m-2}\pi) 
({\lambda{\sqrt V_0}\over
m-2})^{-{2ik\over m-2}} {e^{-ik(\beta +{2\bp\over m-2})}\over \Gamma
(1-{2ik\over m-2})},
$$
The probability for transitions from the classical trajectory
with $\beta=\bp$ to the duality related trajectory $\beta= -
\bp$ is then found to be
\eqn\prob{R_k = {|\Psi_k^{(-)}|^2\over |\Psi_k^{(+)}|^2} =
e^{{-4\pi k\over m-2}}.}
Thus, we see that there is a turning from one branch to another
for this potential where classically forbidden domains exist.
We also notice from \apro\ that the classically forbidden region 
increases with $m$. This is also reflected in \prob\ since $R_k$
increases as $m$ increases.

We note that if we would have chosen $V=-V_0 e^{m\bp}$ instead,
the solutions to the equations of motion along with the constraint
equation lead
$$ \beta ={k\tau\over\lambda},
~~~\bp= {-2\over m-2} {\rm ln} [ {{\lambda\sqrt V_0}\over k}
{\rm cosh}{k(m-2)\tau\over {2\lambda}}].
$$
If we proceed as above the function $\psi_k (\bp)$ would become
a linear combination of the modified Bessel functions $K_\nu
(z)$ and $I_\nu (z)$. But again the requirement of the wave
function to be regular in the limit $\bp\rightarrow \infty$
would fix the wave function to be expressed only in term of
$K_\nu (z)$ and thus asymptotically, in the regime
$\bp\rightarrow -\infty$ , we would have
\eqn\wdwas{\lim_{\bp\rightarrow -\infty} \Psi_k (\bp,\beta) =
\Psi_k^{(+)} + \Psi_k^{(-)}}
where
$$
\Psi_k^{(+)}= - {N\pi\over 2 {\rm sin} ({2ik\pi\over m-2})} ({\lambda
{\sqrt V_0}\over m-2})^{{2ik\over m-2}} {e^{-ik(\beta
-{2\bp\over m-2}})\over \Gamma (1 + {2ik\over m-2})},
$$
and 
$$
\Psi_k^{(-)} = {N\pi\over 2 {\rm sin} ({2ik\pi\over m-2})} ({\lambda
{\sqrt V_0}\over m-2})^{{-2ik\over m-2}} {e^{-ik(\beta +
{2\bp\over m-2}})\over \Gamma (1 - {2ik\over m-2})}.
$$
Thus in this case, in the low energy limit, $R_k\rightarrow 1$
for all $k$. This was expected since in this example the two
branches are smoothly connected at the classical level also. In particular,
in terms of cosmic time $t$, near $t=0$, $\bp$ has an expansion
\eqn\apr{\bp = -{2\over m}{\rm ln}[{\lambda V_0\over k^2}
+ {mk^2\over {4\lambda}}t^2 + ({m\over 4} -1){mk^4\over
{24\lambda^3V_0}}t^4 + ....].}

\bigskip
Now we turn our attention to the second scenario where the
dilaton potential is set to zero and $q\neq 0$. It is evident
from equation \newac\ that the action is no longer invariant under
SFD transformation $\beta\rightarrow -\beta$ and
$\bp\rightarrow \bp$ and if $q$ does not transform nontrivially
under duality. However, we know that the action \oaction\ is
$O(d,d)$ and hence duality transformation invariant. We arrived
at action \newac\ after implimenting Poincare duality \psu\ .
Therefore, in order to maintain duality invariance of action
\newac\ the transformations $(\bp,\beta, q^2) \rightarrow (\bp,
-\beta, q^2 e^{-4{\sqrt 3}\beta})$ should hold good. We can
interprete the presence of the axion term as the effect of a
specific potential which carries both $\beta$ and $\bp$
dependence. It is not possible to obtain classical solutions and
wave function for the WDW equation for a potential with
arbitrary $\beta$
and $\bp$ dependence. We shall see below that, for the problem at
hand, we can define new variables which are linear combinations
of $\bp$ and $\beta$ such that the resulting classical equations
of motion take rather simple form which admit simple solutions
{\foot {Similar transformations were used in the context
of Brans-Dicke theory or various string theories and M theory to reduce
equations of motion of various fields into simple Liouville
and/or Toda like equations~\refs{\pope}.}.
Furthermore, we can solve the WDW equation for this case
{\foot {The equations of motion of string theory in the presence
of the axion field has also been solved in \refs{\clw}, but with
a different parametrization. As we will see, in our parametrization,
it will be easier to analyse the corresponding WDW equation.}}.
Let us
first write the equations of motion in the variables $\beta$ and
$\bp$ in the N=1 gauge. They are
\eqn\qbeta{\partial_\tau^2\beta - {\sqrt 3\over 2} q^2 e^{-2(\sqrt 3\beta 
+ \bp)} = 0,}
\eqn\qphi{\partial_\tau^2\bp + {q^2\over 2} e^{-2(\sqrt 3\beta
+ \bp)} = 0,}
and the constraint is
\eqn\conn{(\partial_\tau\bp)^2 - (\partial_\tau\beta)^2 - {q^2\over 2}
e^{-2(\sqrt 3\beta +\bp)} = 0.}

Defining $X = (\sqrt 3\beta+ \bp)$ and $Y = \beta + \sqrt 3 \bp$, from 
\qbeta\ and \qphi\ we get
\eqn\xe{\partial_\tau^2 X - q^2 e^{-2X} = 0, ~~~\partial_\tau^2 Y =0.}
Now the constraint equation has the following form
\eqn\cons{(\partial_\tau X)^2 -(\partial_\tau Y)^2 + q^2 e^{-2X}
=0.} 
Note that these equations are identical to the corresponding
equations of the earlier example if we identify $X, Y$ and $q^2$ with
$\bp, \beta$ and $-V$ respectively which may have some
interesting significance but we will not proceed futher in this
direction. 
The solutions to the above equations are 
\eqn\valx{X = -{1\over 2}{\rm ln}[{c\over {4q^2}}{\rm sech}^2 {\sqrt 
c\over 2}\ta ],~~~Y = -{\sqrt c \tau\over 2}.}
Using $X$ in \qbeta\ and \qphi\ we get 
\eqn\betao{\beta = {-\sqrt 3\over 4}{\rm ln}[{c\over {4q^2}}{\rm sech}^2
{\sqrt c\tau\over 2}] + {\sqrt c \tau\over 4},}
\eqn\phio{\bp = {1\over 4}{\rm ln}[{c\over {4q^2}}{\rm sech}^2 
{\sqrt 
c\over 2} \tau ] - {{\sqrt {3c}}\tau\over 4}.}
The above solutions are to be used for $\tau > 0 $. For the 
other branch $\tau < 0 $ we have the set of solutions, 
namely,
\eqn\betaoa{\beta = {-\sqrt 3\over 4}{\rm ln}[{c\over {4q^2}}{\rm sech}^2
{\sqrt c\tau\over 2}] - {\sqrt c \tau\over 4},}
\eqn\phioa{\bp = {1\over 4}{\rm ln}[{c\over {4q^2}}{\rm sech}^2 
{\sqrt 
c\over 2} \tau ] + {{\sqrt {3c}}\tau\over 4}.}
We can write down the WDW equation after passing over to the
Hamiltonian description and this is found to be
\eqn\qwdw{[\partial_\bp^2 -\partial_\beta^2 + {1\over 2}
q^2\lambda^2 e^{2(\sqrt 3\beta+ 
\bp)}]\Psi(\beta ,\bp) = 0.}
Note that because of the mixed nature of the last term in the
square bracket we cannot look for solutions by separation of
variables i.e. there is no more conservation of momentum in the
$\beta$-direction. However, equation \qwdw\ can be brought to the
following convenient form for further analysis in terms of $X$
and $Y$ as defined before.
\eqn\nwdw{[\partial_X^2 -\partial_Y^2 + {1\over 2}
q^2\lambda^2 e^{-{2X}}]\Psi (X,Y) = 0.}
It is obvious that we can separate the variables and the wave
function will be factored into an oscillatory part in $Y$ and a
function which is a linear combination of Henkel functions
$H^{(1)}$ and $H^{(2)}$. We also see that $X\rightarrow \infty$,
for $\tau\rightarrow \pm \infty$.
%and it does not go to $-\infty$. 
Thus we can choose $H^{(2)}$ as a factor of the wave
function. 
\eqn\si{\Psi_{k} (X,Y) = N H^{(2)}_{ik}({q\lambda\over \sqrt
2}e^{-X})e^{-ikY}. } 
Now, expanding for $X\rightarrow \infty$ we find that the wave
function can be written as sum of two terms, $\Psi^{(+)} +
\Psi^{(-)}$, as before and in terms of the original variables it
has the following form
\eqn\nsi{\Psi_k (\beta,\bp) = i N {\rm cosec} (ik\pi) [({\lambda q
\over 2{\sqrt 2}})^{-ik} e^{-k\pi} {e^{-ik({\sqrt
3}+1)(\beta+\bp)} \over \Gamma (1-ik)} - ({\lambda q\over
2{\sqrt 2}})^{ik} {e^{ik({\sqrt 3}-1) (\beta -\bp)}\over \Gamma
(1+ik)}].} 
Notice that the above wave function strongly resembeles the
solution obtained in \refs{\gmv}. However, it is worthwhile to
note a few features. First of all, the quantum number $k$
appearing in \refs{\gmv} or the one we introduced in our first
example is the eigen value of the canonical momentum operator in
$\beta$-direction. On the other hand, $k$ appearing here is the
eigen value of $\Pi_Y$, momentum conjugate to $Y$ which is a
linear combination of $\beta$ and $\bp$. Thus, it is not straight
forward to interprete the wave function as linear combination of
left and right moving waves as was presented earlier. We feel it
is quite interesting, in this case, one would solve the
Wheeler-De-Witt equation in the presence of a potential which
depends both on $\beta$ and $\bp$. We must admit that in passing
from classical to quantum hamiltonian, we have not been careful
about the ordering ambiguity. In some cases, however, the $O(d,d)$
symmetry uniquely fixes the ordering (see ref \refs{\gmv} for
a discussion on this issue).
\bigskip

In this note we have considered string effective action in
presence of dilaton 
potential and the axionic field in two separate cases. When the axion is
absent, a generalised form of the dilatonic potential used by
\refs{\gmv} is considered
and the classical solutions were obtained in terms of the dilatonic time,
$\tau$, rather than the cosmic time, t. Then we looked for solutions of the 
Wheeler-De Witt equation for the potential. We chose the boundary condition
such that the intial incoming wave starts as left moving mode and after
scattering off the barrier, we have the right moving mode which corresponds
to the situation of an expanding, decelerating Universe. Subsequently, we
analysed the case, where the dilaton potential was set to zero and the
antisymmetric field strength was included. Since the axionic charge is 
conserved, we have a potential term involving the dilaton in presence of
the axionic charge. The classical equations of motion were solved exactly
after making a suitable coordinate transformation. Subsequently, we solved
the WDW equation for this case and found the appropriate wave function
satisfying the required boundary conditions.

We may remark that in the presence of arbitrary dilaton potential and the
axion, we shall not be able to solve the WDW equation exactly; however, one
could resort to suitable approximation method, such as WKB approximation, as
is utilised in quantum cosmology. Furthermore, it is recognised that the 
dilaton-graviton system has close resemblance with the Brans-Dicke theory
with the specific choice of the parameter. Indeed, recently dualities in
the Brans-Dicke theory have been discussed and the solutions of WDW equation
have been analyzed in this frame work \refs{\lsey}. It will be 
interesting to investigate
the solutions of WDW equations and their properties for axion and dilaton
in such theories.

Finally, we would like to mention that in type II string theories, there
are other antisymmetric rank 3 tensor fields at massless level with 
different dilaton coupling than the one that has been considered
here. Also depending upon the detail of the compactification, there 
can be other moduli fields. It turns out that in many of the cases,
classical equations of motion can be solved exactly
\refs{\ovrut , \pope}. Following closely
the discussion here, it might be possible to solve and analyse the
WDW equations in those cases. We hope to report on it in the future.

{\bf Acknowledgement}: We would like to thank the hospiatality of ICTP 
where the work was initiated. 
The work of SM is supported by NSF Grant No. 
PHY-9411543. \vfill\eject
\listrefs
\bye